\documentclass[sigconf]{acmart}
\usepackage{subcaption}
\usepackage{float}
\usepackage{tabularx}

\AtBeginDocument{%
  \providecommand\BibTeX{{%
    \normalfont B\kern-0.5em{\scshape i\kern-0.25em b}\kern-0.8em\TeX}}}

\setcopyright{acmcopyright}
\copyrightyear{2018}
\acmYear{2018}
\acmDOI{10.1145/1122445.1122456}

\acmConference[Woodstock '18]{Woodstock '18: ACM Symposium on Neural
  Gaze Detection}{June 03--05, 2018}{Woodstock, NY}
\acmBooktitle{Woodstock '18: ACM Symposium on Neural Gaze Detection,
  June 03--05, 2018, Woodstock, NY}
\acmPrice{15.00}
\acmISBN{978-1-4503-9999-9/18/06}

\begin{document}

\title{Leveraging Gene Expression Data and Explainable Machine Learning for Enhanced Early Detection of Type 2 Diabetes}


\author{Aurora Lithe Roy}
\affiliation{%
	\institution{Institute of Engineering \& Management}
	\streetaddress{1 Th{\o}rv{\"a}ld Circle}
	\city{Kolkata}
	\country{India}}
\email{auroralroy2003@gmail.com}

\author{Md Kamrul Siam}
\affiliation{%
    \institution{Colorado School of Mines}
    \city{Golden}
    \state{Colorado}
    \country{USA}
}
\email{siam@mines.edu}

\author{Nuzhat Noor Islam Prova}
\affiliation{%
	\institution{Pace University, Seidenberg School of CSIS}
	\city{One Pace Plaza, New York City}
	\country{USA}}
\email{nuzhatnsu@gmail.com}

\author{Sumaiya Jahan}
\affiliation{%
  \institution{Bangladesh University of Business \& Technology}
  \city{Dhaka}
  \country{Bangladesh}}
\email{sumaiyajahaann@gmail.com}
\author{Abdullah Al Maruf}
\affiliation{%
  \institution{Bangladesh University of Business \& Technology}
  \city{Dhaka}
  \country{Bangladesh}}
\email{marufbubt822@gmail.com}



\begin{abstract}
Diabetes, particularly Type 2 diabetes (T2D), poses a substantial global health burden, compounded by its associated complications such as cardiovascular diseases, kidney failure, and vision impairment. Early detection of T2D is critical for improving healthcare outcomes and optimizing resource allocation. In this study, we address the gap in early T2D detection by leveraging machine learning (ML) techniques on gene expression data obtained from T2D patients. Our primary objective was to enhance the accuracy of early T2D detection through advanced ML methodologies and increase the model's trustworthiness using the explainable artificial intelligence (XAI) technique. Analyzing the biological mechanisms underlying T2D through gene expression datasets represents a novel research frontier, relatively less explored in previous studies. While numerous investigations have focused on utilizing clinical and demographic data for T2D prediction, the integration of molecular insights from gene expression datasets offers a unique and promising avenue for understanding the pathophysiology of the disease. By employing six ML classifiers on data sourced from NCBI's Gene Expression Omnibus (GEO), we observed promising performance across all models. Notably, the XGBoost classifier exhibited the highest accuracy, achieving 97\%. Our study addresses a notable gap in early T2D detection methodologies, emphasizing the importance of leveraging gene expression data and advanced ML techniques. 
\end{abstract}






\maketitle

\section{Introduction}
Diabetes is a chronic metabolic disorder where the human body struggles to regulate blood sugar. More than 537 million individuals were diagnosed with diabetes in 2021, as indicated by recent data from the \href{https://diabetesatlas.org/}{International Diabetes Federation}, and more than 75\% of them lived in low-income and middle-income nations. According to projections, by 2030 and 2045, there will be 643 million and 783 million such cases, respectively. There are three main types: Type 1, Type 2, and Gestational diabetes. Among these, T2D is the most common form. It is a condition caused by either resistance to the effects of insulin or inadequate insulin synthesis by the pancreatic cells. As per an article by the \href{https://www.who.int/news-room/fact-sheets/detail/diabetes}{World Health Organization}, T2D accounts for more than 95\% of diabetes cases.

Preventive measures, including maintaining a healthy weight, regular exercise, and consideration of genetic factors, can often deter the development of T2D. However, symptoms of T2D can initially be subtle and may take years to manifest fully. As a result, diagnosis may be delayed until after the disease has already begun, and in some cases, complications may have already developed. So timely identification of the condition is crucial in averting its severe consequences.

Genetic predispositions affecting insulin production, insulin resistance, and lifestyle choices, including obesity, overeating, and inactivity, are the leading causes of T2D \cite{intro7}. The traditional way of diagnosing T2D is based on clinical symptoms and blood glucose levels, which may not become apparent until the disorder has advanced considerably. As a result, there is an urgent need for more proactive and accurate approaches to diagnose this condition in the early stages. Using ML models enables timely intervention to reduce the risk of complications. 

In T2D, changes in gene expression impact insulin and glucose regulation. When genes involved in insulin production and metabolism undergo alterations, it contributes to insulin resistance and impaired glucose control. Studying these genetic influences is crucial for developing improved treatments and understanding the disease's underlying mechanisms. Gene expression analysis reveals molecular causes and identifies biomarkers in large-scale gene expression data that are not apparent to human observers to assist in the detection of T2D. It reveals dysregulated pathways and pinpoints possible biological targets for more potent remedies. 

Previous research on diabetes prediction has utilized traditional ML techniques like Naïve Bayes, Support Vector Machines, LR, K-nearest Neighbor (KNN) \cite{intro1, intro2, intro3}, as well as deep learning methods such as Artificial Neural Networks \cite{intro3, intro4}. Other methods, such as Sparse Distance Weighted Discrimination, Generalized Additive Models, Generalized Boosted Regression, were also investigated for classifying T2D patients \cite{comparison1}. However, these methods have primarily operated on datasets containing features such as blood glucose levels, blood pressure, and body mass index. Studies have also explored T2D prediction by utilizing gene sequences from genomic DNA fragments to match gene patterns with training data \cite{intro6}. Using ML techniques on gene expression data enables the discovery of subtle molecular patterns linked with early disease stages.

In this research, our objective is to identify genes exhibiting significant expression levels through analysis of RNA sequencing data to differentiate between T2D and non-diabetic individuals based on their respective gene expression patterns. We employed a feature importance technique across six distinct ML algorithms: Decision Tree (DT), Random Forest (RF), Logistic Regression (LR), Gradient Boosting (GB), Extreme Gradient Boosting (XGBoost) and Adaptive Boosting (Adaboost). Evaluating the performance of the models, we found that XGBoost stands out with enhanced prediction accuracy. Furthermore, using XAI enhances the interpretability of the models. This fosters greater confidence in clinical decision-making by shedding light on the predictive mechanisms underlying the models. With this study, we propose a more comprehensive ML-driven approach to understanding the genetic landscape of diseases such as T2D. The contribution of our research :
\begin{itemize}
    \item Our research has explored the complex world of gene expression datasets using cutting-edge computational techniques to identify the genetic signatures underlying T2D.
    \item In the research, six ML classifiers were applied to predict T2D. We have fine-tuned the classifier according to our research. Our proposed model underscores the importance of the ML model in bioinformatics.
    \item The proposed model outperformed the previous research regarding all measurement criteria.
    \item We have applied XAI techniques to enhance the readability and make the model more trustworthy.
\end{itemize}
The research paper is organized as follows: In Section 2, we present the previous studies. In Section 3, we show the proposed model and the dataset description. In Section 4, we show the outcomes of our research.

\section{Related Work}
Several studies have investigated non-genetic variables such as blood sugar level, blood pressure, cholesterol, age, and physical activity status in detecting T2D. Fazakis et al. \cite{lr1} assessed the WeightedVotingLRRFs ensemble model for T2D risk prediction, achieving an AUC of 0.884. Joshi and Dhakal \cite{lr2} identified five major predictors of T2D with 78.26\% accuracy: age, diabetes pedigree function, BMI, glucose, and pregnancy. They also gained a 21.74\% cross-validation error rate.

Karmand et al. \cite{lr4} compared seven ML methods for T2DM prediction in an Iranian setting, with Gradient Boosting Machine outperforming others with an AUC of 0.75 and 0.76 for males and females respectively, and an F1 score of 0.33 for males and 0.42 for females. Furthermore, major features influencing T2DM prediction were identified using SHAP analysis, which included age, blood type, past hospitalization history, and sugar consumption. 

Nipa et al. \cite{lr5} achieved notable accuracies in outlier detection using classifiers like Extra Tree (97.11\%) and Multi-Layer Perceptron (MLP) (96.42\%) and found stable predictive capabilities in methods like Light Gradient Boosting Machine and RF. They used patient data from Bangladesh's Sylhet Diabetes Hospital and the ML repository of the University of California, Irvine. Iparraguirre-Villanueva et al. \cite{lr6} explored KNN (79.6\% accuracy) and Bernoulli Naïve Bayes (77.2\% accuracy) to identify diabetes using the Pima Indian database. Aguilera-Venegas and their team \cite{lr7} predicted T2DM incidence seven and a half years in advance by analyzing 18 socio-demographic and clinical factors, with RF attaining accuracy of 92.91\%.

Many researchers have explored the intersection of bioinformatics and ML to gain insights into the genetic factors underlying T2D. Saxena et al. \cite{lr12} aimed to identify crucial metagenes by wielding LASSO feature selection together with five ML models, including ANN which exhibited excellent performance with a 95\% AUC and achieved 73\% accuracy in differentiating between glucose-tolerance and diabetics. These metagenes were enriched in terms relevant to diabetes along with being identified in earlier genome-wide association studies of T2D. This study helps in understanding insulin resistance and T2D pathways. Xiaoting Pei et al. \cite{lr11} used LASSO regression and RNA sequencing in mice lacrimal glands to identify marker genes associated with T2DM. Wild-type mice were used as controls, and twenty db/db (diabetic) mice were used as T2DM models. Five marker genes were chosen using LASSO regression out of 689 differentially expressed genes. Among these, Synm was downregulated, while Ptprt, Glcci1, Tnks, Elovl6 were upregulated in T2DM mice. This model demonstrated high predictive accuracy with 1.000 and 0.980 AUC respectively in the training and testing datasets. Nevertheless, it has several drawbacks, including a limited sample size that can restrict applicability and a focus on mouse lacrimal glands that might not apply to humans. 

In a retrospective study by Amos Otieno Olwendo et al. \cite{lr14}, electronic health record (EHR) data from Nairobi Hospital were analyzed to identify confirmed cases of diabetes. Supervised and unsupervised learning algorithms were applied. RF demonstrated the highest accuracy at 95\%. GB and MLP yielded 94\% accuracy. Rai et al. \cite{lr13} employed single-nuclei chromatin accessibility profiling to delineate cell-type-specific regulatory signatures in pancreatic islets using the sci-ATAC-seq protocol. They analyzed 1,456 nuclei, identifying major cell-type populations and used a U-Net-based deep learning approach to overcome data analysis challenges. Integrating their results with T2D GWAS variants unveiled intricate regulatory landscapes in T2D pathophysiology, shedding light on cell-specific mechanisms. However, limitations include reliance on a single islet sample.

Middha and Mittal \cite{lr15} presented a novel approach for T2DM detection using Competitive Multi-Verse Rider Optimizer. Their method integrated two classifiers: Rider-based Neural Network, Deep Residual Network. They evaluated various feature selection techniques, among which the Tanimoto similarity approach emerged as the most effective. It achieved 0.932 testing accuracy, 0.932 sensitivity, and 0.914 specificity. 
In reviewing existing studies, it becomes evident that:
\begin{itemize}
    \item Most studies rely on clinical and demographic features data \cite{lr1, lr4, lr5}. However, we choose to work with genetic data because it holds promise for early detection of T2D due to its stability and lower susceptibility to environmental influences.
    \item While numerous existing models exhibit high accuracy, there remains room for improvement as accurate diagnosis holds paramount importance in the medical domain \cite{lr2, lr6, lr7, lr12, lr4}. Our model showcases markedly superior accuracy compared to established models targeting human genes linked to T2D.
    \item The prevalent utilization of unbalanced datasets often results in low accuracy rates \cite{lr1, lr4, lr6}. This was another recurring observation. For the minority class, the result was biased. Despite operating with such a dataset, our proposed model showcases notably robust performance.
\end{itemize}

\section{Proposed Model}
\subsection{Dataset Description and Preprocessing }
We used the GSE81608 \cite{data1} dataset from the Gene Expression Omnibus (GEO) database \cite{data2}. This dataset contains RNA sequencing data derived from a study on 1600 human pancreatic islet cells, encompassing various cell types including $\alpha$, $\beta$, $\delta$, and Pancreatic Polypeptide (PP) cells. These cells were obtained from both non-diabetic and Type 2 Diabetes (T2D) organ donors. This data was generated using the Illumina HiSeq 2500 as the sequencing platform. There were 949 T2D samples and 651 non-diabetic samples. Each sample in the dataset quantifies expression levels for 28089 genes. The single-cell RNA sequencing technique helps to see how gene expression varies between different cell types within the islet and how gene expression differs in healthy individuals compared to those with T2D. We downloaded both the normalized gene-level data and the complete metadata from \href{http://www.ilincs.org/apps/grein/?gse=GSE81608}{GREIN} for easy access and analysis \cite{data3}. The metadata contains sample details, such as condition, age, ethnicity, gender, and cell subtype.
  
The dataset obtained from GREIN was preprocessed using the PANDAS library. Techniques such as transposition and modification of column headers were applied to improve dataset organization. Then, the column containing sample conditions (Type 2 Diabetes or non-diabetic) from the metadata was merged with the gene expression data based on sample names to integrate sample conditions with gene expression profiles. Finally, the condition labels were converted into binary format. '1' was designated to samples with T2D and '0' to non-diabetic samples. We have used the train\_test\_split method in our research to divide the dataset into training and testing subsets. 20\% data was utilized for testing, while the remaining data was used for training.

\subsection{Classification Algorithms}
In this study, our objective was to use a variety of commonly used ML models to obtain an optimized performance. We aimed to explore decision-based, ensemble, and linear models to assess their efficiency. As such, we utilized six models for classification: DT, RF, LR, AdaBoost, GB, and XGBoost. Each method has a unique strength: AdaBoost highlights misclassified cases, GB optimizes performance through gradient descent, XGBoost iteratively increases performance, DT visualizes decision boundaries, RF combines tree predictions, and LR calculates probabilities. We have used different parameters and values for the classifier. Table \ref{tabc} shows the classifiers' parameters and values.\\\\

\textbf{Decision Trees: }
Across regression and classification applications, DTs are flexible and comprehensible ML instruments. Specifically in cases of binary classification, DTs work by repeatedly dividing the feature space according to feature thresholds. Decision nodes that match anticipated class outcomes are produced by this procedure, providing a clear understanding of the underlying patterns in the data. A particular feature is assessed at each decision node, directing the algorithm along branches according to the feature's result. DTs are known for their simplicity and are excellent at illustrating intricate linkages in the data. Let $T$ be the DT model. At each decision node $i$, the algorithm selects a feature $f_i$ and a threshold $\theta_i$. The decision rule at node $i$ can be represented as:
\[
\text{if } x_{f_i} \leq \theta_i \text{ then } T(x) = T_L \text{ else } T(x) = T_R
\]
Where:
\begin{itemize}
    \item $x_{f_i}$ is the value of feature $f_i$ for the input $x$,
    \item $\theta_i$ is the threshold value,
    \item $T_L$ and $T_R$ are the left and right child trees, respectively.
\end{itemize}

This process recursively runs until a stopping condition is satisfied, such as reaching a tree depth of maximum value or no further improvement in classification performance.\\

\textbf{Random Forest: }
RF is an ensemble learning method that combines multiple DTs to improve predictive accuracy and reduce overfitting. Mathematically, let $T_1, T_2, ..., T_n$ represent the individual DTs in the RF ensemble. A random portion of the training data and a random subset of the features are used to train every $T_i$. The prediction of the RF ensemble $\hat{y}_{\text{RF}}$ for a given input $x$ is typically obtained by aggregating the predictions of individual trees. 

In classification tasks, a common aggregation method is the majority vote, where the final predicted class $\hat{y}_{\text{RF}}$ is determined by the class that receives the most votes from the individual trees:

\[
\hat{y}_{\text{RF}}(x) = \text{argmax}_{y} \sum_{i=1}^{n} \mathbb{1}(T_i(x) = y)
\]

where $\mathbb{1}(\cdot)$ is the indicator function, and $T_i(x)$ represents the prediction of the $i$-th DT.

RF introduces randomness at two levels:
\begin{enumerate}
    \item \textbf{Bagging (Bootstrap Aggregating)}: A random sample of training data with replacement is used to train every tree in the forest. This random sampling contributes to the trees' increased diversity.
    \item \textbf{Feature Randomness}: A random subset of features is taken into account at each split in the tree to determine the optimal split. This increases the diversity of the trees even further.
\end{enumerate}

The model's generalizability is enhanced and overfitting is lessened by the randomness of the feature selection process and the data. Additionally, aggregating the predictions of multiple trees helps reduce variance and provide more stable predictions.\\

\textbf{Logistic Regression: }
One popular statistical method for binary classification problems is LR. It uses the logistic function to establish a relationship between one or more independent variables and a dependent binary variable. LR calculates probabilities using this link, producing a probability score between 0 and 1. This number represents the probability that a certain occurrence is a member of the positive class. For this reason, LR is a useful technique for identifying binary outcomes and measuring the degree of ambiguity surrounding classification judgments.\\

\textbf{Adaboost: }
AdaBoost is an ensemble learning technique that iteratively combines weak classifiers to create a strong classifier. Let's denote the weak classifiers as $h_1(x), h_2(x), ..., h_T(x)$, where every $h_t(x)$ focuses on the prior iterations' incorrectly identified cases, and is trained those portions of the training data. The final strong classifier $H(x)$ is a weighted combination of these weak classifiers:

\[
H(x) = \text{sign}\left( \sum_{t=1}^{T} \alpha_t h_t(x) \right)
\]

Where:
\begin{itemize}
    \item $\alpha_t$ is the weight assigned to the $t$-th weak classifier, determined by its performance in classifying the training data,
    \item $\text{sign}(\cdot)$ is the sign function, which converts the weighted sum into a binary classification decision.
\end{itemize}

The training process of AdaBoost can be described as follows:
\begin{enumerate}
    \item Initialize the weights for each training instance equally.
    \item For $t = 1$ to $T$:
    \begin{itemize}
        \item Train a weak classifier $h_t(x)$ using the weighted training data.
        \item Compute the error $\varepsilon_t$ of the weak classifier, which is the weighted sum of misclassified instances.
        \item Compute the weight $\alpha_t$ of the weak classifier, which is proportional to its classification accuracy.
        \item Update the weights of the training instances, giving higher weights to misclassified instances.
    \end{itemize}
    \item Repeat until $T$ weak classifiers are trained.
\end{enumerate}

AdaBoost emphasizes the significance of misclassified instances by assigning them higher weights in subsequent iterations. This iterative process ensures that AdaBoost focuses on difficult-to-classify instances, leading to a robust and accurate ensemble model.\\

\textbf{Gradient Boosting: }
GB is a boosting technique within a functional space that combines weak learners sequentially to correct errors. Unlike traditional boosting, GB focuses on pseudo-residuals instead of conventional residuals.

GB computes the error by comparing predicted and actual values and aims to minimize a designated loss function (e.g., cross-entropy or mean squared error). The model continually modifies its architecture throughout training in order to reduce this loss. The objective function combines the loss function's sum across all instances with a regularization term penalizing weak learner complexity. The residuals are the difference between the true values and the ensemble's predictions up to the $(n-1)$-th tree for each instance. The residual $r_{kn}$ is computed as
\begin{equation}
r_{kn} = -\left[ \frac{\partial L(y_k, \hat{y}_k)}{\partial \hat{y}_k} \right]_{\hat{y}_k = f_{n-1}(x_k)}
\end{equation}
where $f_{n-1}(x_k)$ denotes the prediction of the ensemble up to the $(n-1)$-th tree for the $k$-th instance. $L(y_k, \hat{y}_k)$ represents the loss function measuring the discrepancy between true and predicted values.

The update rule for the next weak learner incorporates the learning rate (\( \theta \)) and the weak learner (\( w_n(x) \)) to be fitted in the \( n \)-th iteration. It is defined as:
\begin{equation}
f_n(x) = f_{n-1}(x) + \theta w_n(x)
\end{equation}
where \( f_{n-1}(x) \) is the prediction of the ensemble up to the \( (n-1) \)-th tree.

Following these operations, GB iteratively refines the ensemble to minimize the loss function and enhance predictive performance while mitigating overfitting risks across training, validation, and test datasets.\\

\textbf{XGBoost: }
An ensemble learning technique called XGBoost uses a succession of weak learners, usually DTs, to increase prediction performance. The algorithm operates by iteratively adding new models to fix the mistakes caused by the prior ones.

Gradient descent is used to minimize a loss function, which gauges the discrepancy between expected and actual values, in order to optimize XGBoost. The representation of the XGBoost objective function is as follows: 
\begin{equation}
\text{Objective} = \sum_{i=1}^{n} \text{Loss}(\hat{y}^i, y^i) + \sum_{k=1}^{K} \Omega(f^k)
\end{equation}
where:
\begin{itemize}
    \item $\hat{y}^i$ represents predicted value for the $i$-th observation,
    \item $y^i$ represents actual value for the $i$-th observation,
    \item $\Omega(f^k)$ is the regularization term applied to the $k$-th weak learner $f^k$,
    \item $K$ is the total number of weak learners.
\end{itemize}

XGBoost introduces new weak learners stepwise throughout training, with each learner concentrating on the ensemble's incorrect predictions. The ultimate prediction results from combining the forecasts of all weak learners.

\begin{figure}[!htbp]
    \includegraphics[width=3.6 in]{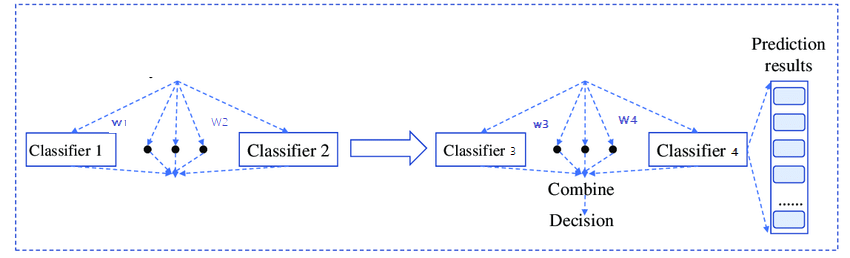}
    \caption{General Architecture of XGBoost}
    \label{xgb}
\end{figure}

The algorithm uses gradient descent to optimize the objective function; gradients for the loss function and the regularization term are computed. XGBoost seeks to minimize the total objective function by iteratively updating the model parameters, producing a highly accurate predictive model. The general architecture of the XGBoost classifier is shown in Figure \ref{xgb}.

\begin{table}[htbp]
\caption{Classifier Parameters and Values}
\centering
\begin{tabular}{|l|c|c|c|c|}
\toprule
\textbf{Classifier} & \textbf{Parameters} & \textbf{Values} \\
\hline
Decision Tree (DT) & Max Depth & 10 \\
                   & Min Samples Split & 2 \\
                   & Criterion & Gini \\
\hline
Random Forest (RF) & N Estimators & 100 \\
                   & Max Depth & 20 \\
                   & Min Samples Split & 5 \\

\hline
Gradient Boosting (GB) & Learning Rate & 0.1 \\
                    & N Estimators & 100 \\
                    & Max Depth & 3 \\
                    & Subsample & 1.0 \\
                    & Min Samples Split & 2 \\
                
\hline
XGBoost & Learning Rate & 0.1 \\
              & Max Depth & 6 \\
              & Subsample & 0.8 \\
              & Colsample Bytree & 0.8 \\
              & Gamma & 0 \\
              & Lambda & 1 \\
              & Alpha & 0 \\
              
\hline
AdaBoost & N Estimators & 50 \\
                     & Learning Rate & 1.0 \\
                     & Base Estimator & Decision Tree \\
\hline
Logistic Regression (LR) & Penalty & L2 \\
                          & C & 1.0 \\
                          & Solver & lbfgs \\
\bottomrule
\end{tabular}
\label{tabc}
\end{table}

\section{Results and Analysis}
We carefully examined the research findings in this section. We first evaluated the models' capacity for generalization before delving into the complex worlds of overfitting and underfitting. Using a careful combination of training and testing accuracy, we measured how likely each model was to fall into these traps. After that, we looked at more than just accuracy metrics. We started a thorough investigation, examining the models from the perspective of true positives and negatives compared to false positives and negatives. This analysis made it easier to compute recall, precision, and their harmonic mean, or F-score.
To find the best model, we used a variety of criteria to identify the best candidates. We also explored the depths of AUC ROC and Confusion Matrix analyses rather than focusing only on accuracy to make our decision. Though we were aware of the inherent imbalance in our dataset, we went further. We also included MCC and Kappa values to strengthen our analysis and ensure that the top model was confidently chosen.

\subsection{ Model Generalization Analysis}
A detailed snapshot of the testing and training accuracies, as well as the corresponding execution times for each model, are shown in Table \ref{acc}. This multifaceted dataset forms the basis of our analysis, providing insight into the inherent complexity involved in the models' execution and their effectiveness in classifying diabetics.


\begin{table}[htbp]
  \caption{Accuracy scores on test data and train data, and execution times of the ML models}
  \label{acc}
  \centering
  \begin{tabular}{lccc}
    \toprule
    \textbf{Model} & \textbf{Testing Acc.} & \textbf{Training Acc.} & \textbf{Execution Time} \\
    \midrule
    XGBoost & 97\% & 100\% & 3min 36s \\
    RF & 75\% & 100\% & 7.8s \\
    DT & 87\% & 100\% & 13.3s \\
    LR & 85\% & 100\% & 2min 23s \\
    AdaBoost & 94\% & 99.53\% & 1min 25s \\
    GB & 97\% & 100\% & 6min 23s \\
    \bottomrule
  \end{tabular}
\end{table}

From Table \ref{acc}, the comparison of accuracy scores of models on the test dataset reveals that predictions of XGBoost and GB were superior. While all three boosting algorithms, including AdaBoost, achieved accuracies > 90\%, XGBoost and GB outperformed every model with 97\% accuracy. AdaBoost fell slightly behind, with an accuracy of 94\%. Conversely, RF, DT, and LR yielded lower testing accuracies of 75\%, 87\%, and 85\%, respectively.

Besides, AdaBoost achieved a training accuracy of 99.53\%, while all other models attained a perfect training accuracy of 100\%. RF, DT, and LR had much lower testing accuracies compared to their training accuracies, with differences of 25\%, 15\%, and 13\%, respectively. The analysis revealed that the poor-performing classifier encountered challenges with overfitting, resulting in suboptimal model generalization. However, given the very high testing accuracies of the three boosting algorithms, there appears to be no significant overfitting issue. AdaBoost demonstrated a difference of 5.53\% between its training and testing accuracies, while XGBoost and GB exhibited the least overfitting with a difference of only 3\%.

\begin{figure}[!htbp]
  \centering
    \includegraphics[width=2 in]{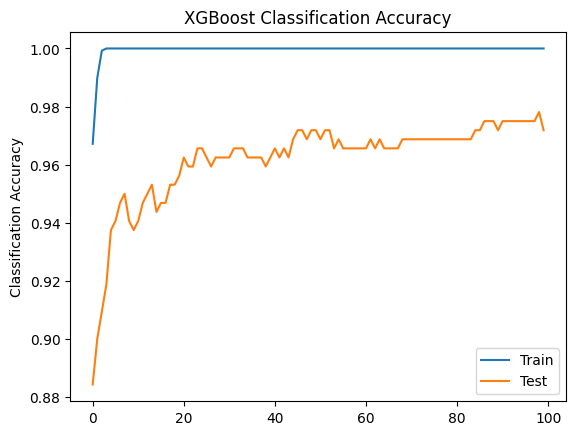}
    \caption{Training and validation accuracy of XGBoost model}
    \label{fig:sub1}
    \end{figure}

\begin{figure}[!htbp]
    \includegraphics[width=2 in]{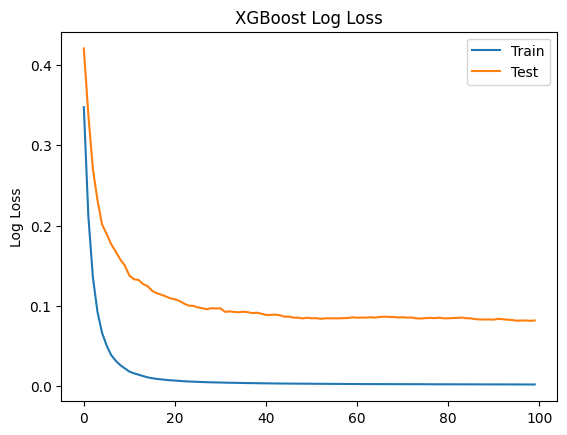}
    \caption{Training and validation loss of XGBoost model}
    \label{fig:sub2}
\end{figure}

However, although XGBoost and GB achieved the highest accuracy, they also required more execution time than other models. XGBoost took 3 minutes 36 seconds, while GB had the longest execution time of all models, 6 minutes 23 seconds. This indicates that these two models are computationally more intensive than the others. XGBoost is the model that performs the best at generalization out of all the ones that were evaluated while overfitting presented difficulties for several classifiers. The training and validation losses and accuracies of our top model is shown in figures \ref{fig:sub1} and \ref{fig:sub2}.


\subsection{Analysis of Classification Reports}
\begin{table}[htbp]
  \caption{Precision, Recall and F-score values of all classifiers}
  \label{metrics}
  \centering
  \begin{tabular}{lccc}
    \toprule
    \textbf{Model} & \textbf{Precision (\%)} & \textbf{Recall (\%)} & \textbf{F1 Score (\%)} \\
    \midrule
    XGBoost & 97.91 & 96.90 & 97.40 \\
    RF & 78.37 & 89.69 & 83.65 \\
    DT & 88.20 & 88.65 & 88.43 \\
    LR & 92.77 & 86.08 & 89.30 \\
    AdaBoost & 94.79 & 93.81 & 94.30 \\
    GB & 96.89 & 96.39 & 96.64 \\
    \bottomrule
  \end{tabular}
\end{table}
 We also assessed the recall, precision, and F1 score metrics in detail. Precision measures how well the model predicts positive outcomes. Out of all occurrences predicted to be positive, it shows the proportion of accurately predicted positive cases. With a minimal false positive rate, XGBoost's excellent precision of 97.91\% indicates that it is capable of correctly recognising instances of diabetes. Table \ref{metrics} displays precision, recall, and F1 scores for the six implemented models.

It can be seen that XGBoost achieved the highest scores across all metrics. Its precision score of 97.91\% signifies that nearly all predicted T2D cases were true positives. Additionally, the recall score indicates that 96.90\% of actual positive cases were correctly identified as T2D. With an F1 score of 97.40\%, XGBoost demonstrates a well-balanced performance between precision and recall that reflects its effectiveness in accurately classifying both T2D and non-diabetic cases. Recall, sometimes referred to as sensitivity, gauges how well the model can accurately identify every positive instance. It represents the percentage of real positive cases that the model accurately detects. With recall rates of 93.81\% and 96.90\%, respectively, XGBoost and AdaBoost show that they are adept at correctly identifying diabetic cases. The F1 score fairly evaluates the model's performance by combining precision and recall into a single metric. In cases where the dataset is unbalanced, it is beneficial. With F1 scores of 96.44\% and 97.40\%, respectively, XGBoost and GB demonstrate their balanced performance in recall and precision. Although its precision is not as high as that of other models, RF shows excellent recall, indicating that it can accurately identify cases of diabetes but may produce more false positives. Although they perform marginally worse than boosting algorithms like XGBoost and AdaBoost, DT and LR demonstrate competitive precision, recall, and F1 scores.


\begin{table}[htbp]
\centering
\caption{Kappa and MCC values}
\label{tab:performance}
\begin{tabular}{@{}lcc@{}}
\toprule
\textbf{Model}          & \textbf{Cohen's Kappa} & \textbf{MCC} \\ \midrule
XGBoost                  & 94.27\%                         & 94.28\%                                           \\
GB        & 93.74\%                         & 93.06\%                                           \\ \bottomrule
\end{tabular}
\end{table}

\subsection{Agreement and Error Analysis}
Given the similarity in accuracy scores between GB and XGBoost, we compared these two models using Cohen's Kappa statistic and Matthews Correlation Coefficient (MCC). 

In terms of Cohen's Kappa, XGBoost attained a value of 94.27\%. GB achieved a slightly lower value of 93.74\%. Both values indicate a high level of agreement between the predicted classifications and the ground truth labels. However, XGBoost outperformed GB by 0.53\%.

The Matthews Correlation Coefficient (MCC) considers true positives, false positives, and false negatives. XGBoost achieved an MCC value of 94.29\%, while GB achieved a slightly lower MCC value of 93.81\%. XGBoost's higher MCC value reinforces its superiority over GB in predictive performance.

\begin{figure}[!h]
    \includegraphics[width=3 in]{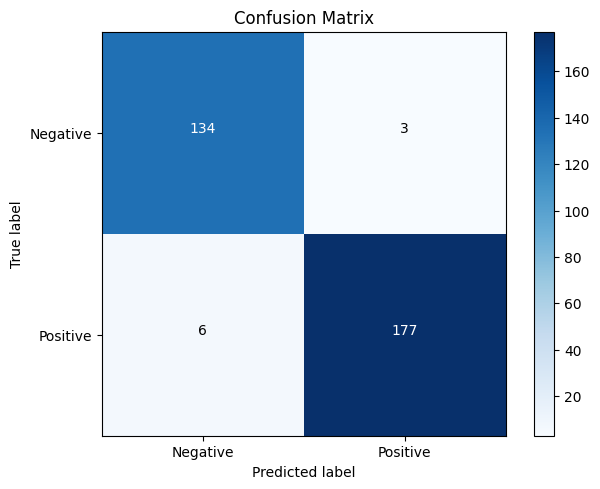}
    \caption{Confusion matrix  of XGBoost model}
    \label{cm}
\end{figure}

XGBoost is the best performer in our study, demonstrating its superior predictive accuracy. A confusion matrix shows how true and predicted classifications interact in Figure \ref{cm}. The confusion matrix, which captures the complex balance between actual and expected outcomes, is a key performance evaluation tool in the classification models. True Positives (TP) are those cases in which the model agrees with the real positive cases and correctly predicts positive outcomes. True Negatives (TN) represent the model's accurate detection of negative results, reflecting the real negative occurrences. False Positives (FP), or Type I Errors, happen when the model predicts positive results even though the real values are negative. False Negatives (FN), also known as Type II Errors, on the other hand, occur when the model misidentifies positive outcomes and incorrectly labels them as negative. It is essential to comprehend these matrix elements to assess a model's predictive performance. As we evaluated the performance of our model, we found that it was quite accurate in classifying positive instances—177 cases were correctly classified as True Positives or TP. Similarly, the model successfully classified 134 data points (True Negatives, TN), indicating that it could identify negative instances. Nonetheless, there were instances of misclassification: six negative class data were mistakenly labeled as positive (False Positives, FP), while three positive class data were incorrectly categorized as negative (False Negatives, FN). These metrics summarize the model's advantages and potential improvement areas, providing vital information for additional optimization.

\subsection{Discussion and Analysis}
\begin{figure}[!h]
    \includegraphics[width=2 in]{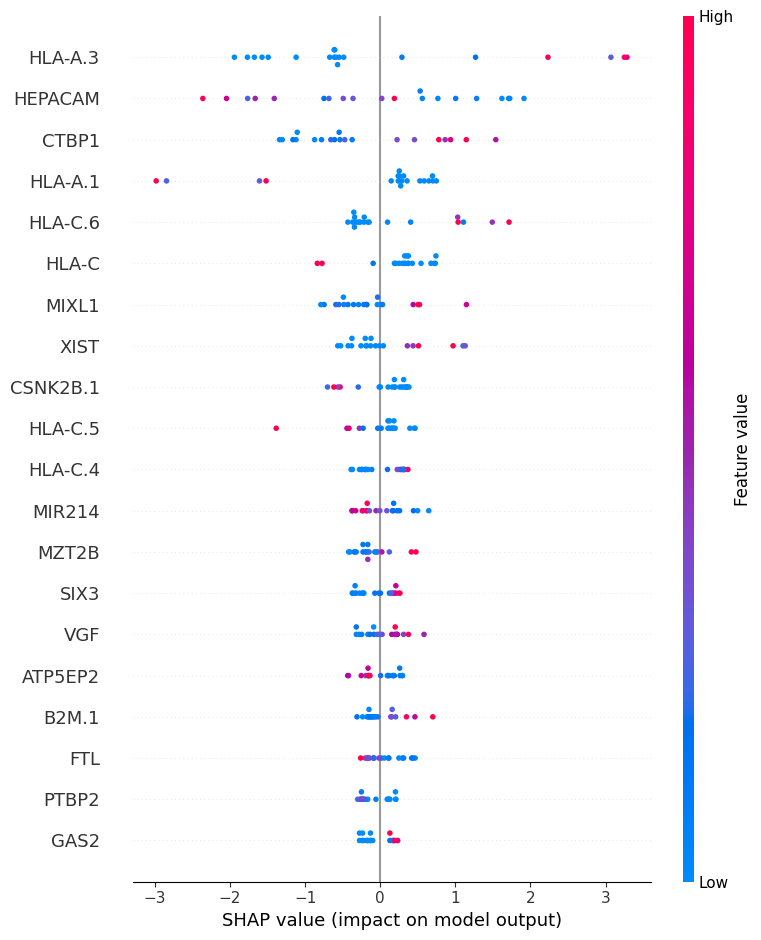}
    \caption{SHAP features impact on the model's prediction}
    \label{shap}
\end{figure}

In this study, we delve into the realm of XAI by examining SHAP (SHapley Additive exPlanations) values, a popular method for explaining individual predictions of ML models. Specifically, we focus on analyzing the SHAP values derived from a subset of the test data comprising the first 20 samples.
We examined the SHAP values for X\_test[0:20], a subset of the test data. For individual predictions, SHAP values explain how each feature affects the model's output. The test set's first 20 samples' SHAP values are computed in this instance. This produces a summary plot of the previously determined SHAP values. The summary plot gives a general idea of the features' relative importance by showing the distribution of SHAP values for each feature across the samples. It aids in determining which features—and in what direction—are most important to the model's predictions.
In the SHAP summary plot (figure \ref{shap}), features' effects on the model's predictions are indicated by color coding: red indicates a positive impact, while blue signifies a negative effect. The features are arranged from highest to lowest importance, with "HLA-A.3" being the most important and "GAS2" being the least important.

The presence of red and blue colors for a single feature suggests that its influence on the model's predictions varies across different samples in the dataset. Higher feature values positively contribute to the model's output for some samples, resulting in red-coloured SHAP values. However, for other samples, the output of the model decreases with higher values of the same feature, represented by blue-coloured SHAP values. This variability often arises due to the nonlinear or complex relationship between the feature and the target variable.

The SHAP summary plot provides a deep understanding of how each feature impacts the model's predictions, highlighting the dynamic nature of these relationships across different data instances.

\begin{table}[htbp]
  \caption{Comparative analysis of our research }
  \label{study_comp}
  \centering
  \begin{tabular}{|c|p{2.5cm}|p{2.5cm}|c|}
    \hline
    \textbf{Ref.} & \textbf{Approach} & \textbf{Dataset} & \textbf{Accuracy} \\
    \hline
    \cite{lr3} & Neural Network ensemble & NHANES and MIMIC & 86\%  \\
    \hline
    \cite{comparison1} & Generalized Boosted Regression Model &  Pima Indian  & 90.91\% \\
    \hline
    \cite{comparison2} & GB & Survey data & 96.90\% \\
    \hline
    \cite{intro3} & RF & Administrative healthcare data  & 84.95\% \\
    \hline
    \cite{intro6} & RNN & miRbase-18.0 & around 82\% \\
    \hline
    \cite{comparison3} & MLP & GEO (GSE22309) & 95\% \\
    \hline
    proposed & XGBoost & GEO (GSE8160) & 97\%\\\hline
  \end{tabular}
\end{table}
We have compared our findings with several other noteworthy studies to place our work in the larger context of diabetic prediction studies (Table \ref{study_comp}). Using a neural network model, Agliata et al. \cite{lr3} obtained an accuracy of 86\%. They were interested in people's health and risk factors, including blood pressure, age, gender, weight, cholesterol, and blood pressure. In a different study, researchers \cite{comparison1} used traditional contextual datasets that included demographic, physiological, and clinical data to apply boosted regression on the Pima Indian Diabetes database. They achieved an accuracy of 90.91\%. With the use of conventional features and a survey, Ganie et al. \cite{comparison2} were able to predict diabetes with a remarkable accuracy of 96.90\% using a GB model. Using administrative healthcare claim data from the CBHS dataset, Lu et al. \cite{intro3} employed a RF classifier and achieved an accuracy of 84.95\%. On the other hand, research on gene expression datasets was done by Srinivasu et al. \cite{intro6} and \cite{comparison3}, who obtained 82\% and 95\% accuracy rates, respectively. 
In our study, we harnessed the power of biological mechanism datasets to forecast the prevalence of diabetes. Our proposed model outperformed previous studies across classical and gene expression datasets. This achievement underscores the efficacy of integrating molecular insights into predictive modeling, paving the way for more accurate and nuanced approaches to diabetes prediction.

\section{Conclusion}
Research on T2D diagnosis and management has gained significant attention recently. Our research proposes a framework for T2D detection utilizing gene expression level data through ML techniques. We have used RNA sequencing data from 1600 human pancreatic islet cells sourced from the GEO database. Our findings from a rigorous evaluation of various ML models demonstrate that XGBoost performs better than the other models in terms of several performance evaluation metrics, with 97\% accuracy. The suggested model also performs better than previous studies in ML-based diabetes diagnosis. This research highlights the potential of ML in early detection of diseases based on gene expression data. Identification of T2D markers even before full symptom manifestation allows the beginning of treatment and lifestyle adjustments within the early stages of diabetes. It paves the way for improved early diagnosis and intervention strategies for diabetes which can significantly improve patient outcomes, delay disease development, and minimize the likelihood of health complications. In future research, it is imperative to broaden model validation by incorporating more datasets to reveal additional diabetes risk factors. Applying deep learning-based pattern recognition models to analyze gene sequences for predicting future illnesses is another promising avenue. Furthermore, another avenue worth exploring is integrating techniques to predict future glucose levels.

\bibliographystyle{ACM-Reference-Format}
\bibliography{main.bib}

\appendix

\end{document}